# A DIPOLE VORTEX MODEL OF OBSCURING TORI IN ACTIVE GALAXY NUCLEI

E.Yu. Bannikova, V.M. Kontorovich

*Institute of Radio Astronomy NAS of Ukraine, Kharkov, Ukraine,*
*Karazin Kharkov National University, Kharkov, Ukraine*
E-mail: *bannikova@astron.kharkov.ua* , *vkont@ira.kharkov.ua*

**Abstract –** The torus concept as an essential structural component of active galactic nuclei (AGN) is generally accepted. Here, the situation is discussed when the torus "twisting" by the radiation or wind transforms it into a dipole toroidal vortex which in turn can be a source of matter replenishing the accretion disk. Thus emerging instability which can be responsible for quasar radiation flares accompanied by matter outbursts is also discussed. The "Matreshka" scheme for an obscuring vortex torus structure capable of explaining the AGN variability and evolution is proposed. The model parameters estimated numerically for the luminosity close to the Eddington limit agree well with the observations.

## 1. Introduction

Starting with the Antonucci and Miller notable work [1], a torus has been considered as an AGN-structure's necessary element forming the basis of the AGN unified model [2, 3]. A brilliant achievement was the first direct observation of obscuring tori described by Jaffe and his colleagues [4] (see also references to recent observations and discussion in [5]). Tori were positively confirmed existing when they had been observed with the MIDI IR-camera equipped VLT optical interferometer, though the efforts to reveal their structure detail and internal motion are yet to come. Many papers are dedicated to tori as embodiments of thick accretion disks, also investigating the stability of these latter defined by orbital motion gradients [6, 7]. However, within the AGN structure, they are mainly considered purely geometrically.

We offer to consider the torus as a dynamic object with its proper vortex motion[1]. As is well known, a torus allows two independent rotations: "orbital" over its periphery and "vortical" (here quoted terms relating to torus motion are ours) over its inner-radius circle. This latter will be of our major interest.

---

[1] Monographs [8, 9 and 10] are dedicated to modern discussion of the theory of vortices. The orbital motion of a self-gravitating torus was investigated in an ample quantity of works since those classical of A. Poincare and S. Kovalevskaya [11], mainly, in view of the problem of Saturn's rings, see later discussions and references in [12].

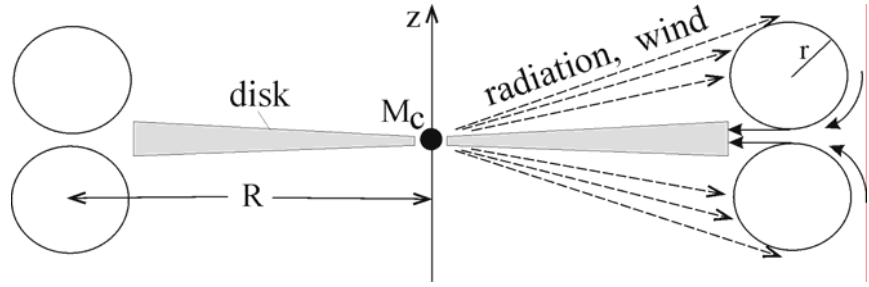

Fig. 1a. Dipole toroidal vortex in the AGN center: an orthogonal-to-disk symmetry plane section, z being the axis of symmetry. Arrows show the possible matter motion directions.

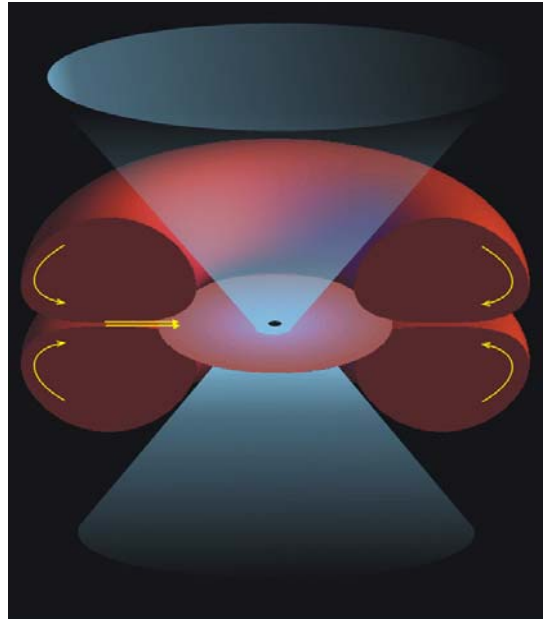

Fig. 1b. Dipole toroidal vortex in the AGN center: 3D picture.
Cones sketch out the wind and radiation.

The vortical motion in a self-gravitating torus (see discussion in [13]) is essentially different from the orbital one, which in an oversimplified case merely means rotation of a torus as a single whole about the axis of symmetry. For the luminosity close to the Eddington limit $L \approx L_{Edd}$, when the gravitation is largely compensated by light pressure, this type motion in the AGN is not so much essential. Though it is necessary to stabilize the self-gravitation of a compact toroidal vortex [14], as it was used there, at first the orbital motion can be well neglected. The vortical torus motion, which as a matter of fact forms a vortex torus, will be of most importance in the following. Originating or being sustained by radiation of the central source or wind "twisting", it is capable of "replenishing" the accretion disk mass, thereby adjusting the process of accretion and introducing a feedback (Fig. 1). Here, the dipole structure of a toroidal vortex which is defined by the symmetry of radial-outflowing wind and radiation is of importance. Note that the streamline structure across such a dipole vortex resembles the structure and topology of streamlines in the well-studied hydrodynamic models, such as Hill and Lamb's vortices [11, 15], Larichev-Reznik solitons [16], and others. At the same time, each component of a toroidal dipole taken separately resembles the Maxwell vortex [15], though counterrotating.

## 2. Vortex twisting by the radiation

At the distance of a torus major radius $R$, light pressure emitted by the central source is $L/(4\pi cR^2)$. The equation for a vortical motion momentum takes the form:

$$\left(\frac{dp_\varphi}{dt}\right)_{twist} = \frac{L}{4\pi cR^2} 2\pi R \cdot \pi r \cdot r \cdot \varsigma(\theta), \qquad (2.1)$$

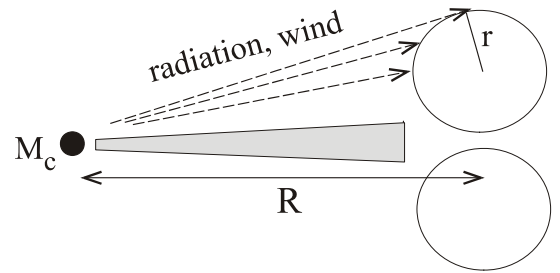

where the right-hand side is the modulus of a force twisting moment with the arm of about a torus minor radius $r$, which appears due to the radiation pressure on the inner torus surface ($\approx 2\pi R \cdot \pi r$). Factor $\varsigma(\theta) \leq 1$ takes into account the torus shape effect and the radiant flux angular dependence[2]. The momentum [13] is related with circulation and mass as $p_\varphi = M_{ring}\Gamma/2\pi$, where $M_{ring}$ is the torus mass and $\Gamma = \oint \mathbf{v}d\mathbf{r} = 2\pi r \cdot v_\varphi$ is the velocity circulation of the torus inner circle. Twisting, which transforms the torus into a toroidal (ring) vortex and sustains its vortical motion (velocity circulation), by virtue of the symmetry should result in a "dipole" vortex whose "northern" and "southern" components rotate in opposite directions (Fig. 2). The streamline cross-section should resemble a pair of vortices of different signs. Such a system, as is known, moves as a single whole with the velocity $V_{ring} = \Gamma/(4\pi r)$ (see, e.g. [11]). As Lamb notes, this motion (Fig. 3) can be interpreted as the necessity to compensate the attraction of vortices induced by the Bernoulli effect arising due to a flow of a moving vortex pair.

Fig. 2. Motion of a vortical pair in the medium [11].

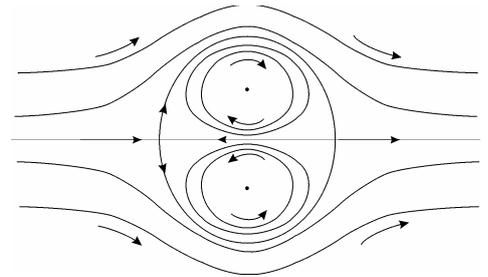

Fig. 3. Scheme of a central source wind- and radiation-twisted vortex.

In our case, such a flow should be due to the central source wind with the velocity $U_{wind}$. As both torus and wind have different densities, the balance condition (see e.g. [17]), as it is easy to

---
[2] By virtue of the possible compensation of the opposite twisting moments applied to different areas of a torus, the coefficient $\varsigma$ may appear much smaller than unit: $\varsigma \ll 1$. It will be noted that its magnitude essentially depends upon the form of a torus section, which in turn must itself be determined with the wind and radiation influence considered.

ascertain, takes the form

$$\rho_{wind} U_{wind}^2 = \rho V_{ring}^2. \qquad (2.2)$$

The fact that both components of a dipole-vortex torus gravitate towards each other should be taken into account too.

Using the known result for the attraction of two electrically charged rings [17], we may rewrite the gravitational force between the two rings with masses $M_a$ and $M_b$ and with the distance $2r$ between them in the form:

$$F_g = \frac{GM_a M_b \cdot 2r \cdot k}{4\pi R^3} \frac{E(k)}{1-k^2},$$

where $k = R/\sqrt{r^2 + R^2}$, $E(k)$ is the elliptic function. If $r \ll R$, then $k \approx 1 - 1/2 \cdot (r/R)^2$, $1 - k^2 \approx (r/R)^2$, and in this case $E(k) \approx E(1) = 1$. The gravitational force between the two components of a dipole toroidal vortex (for $r \ll R$) takes the form

$$F_g = \frac{GM_{ring}^2}{2\pi R r}. \qquad (2.3)$$

This attraction will also be balanced by wind flow. Therefore (see Appendix A), in the equality (2.2) an additional addend appears:

$$\rho_{wind} U_{wind}^2 = \rho V_{ring}^2 + \rho V_{esc}^2, \qquad (2.4)$$

where $V_{esc}^2 = GM_{ring}/(2R)$. Below it will be shown that for the numerical parameters chosen, the gravity contribution (i.e. the second addend in the right-hand side of equality (2.4)) exceeds the hydrodynamic one and is about the same order as the contribution of radiation twisting. Therefore in this work, the wind effect is neglected. At the same time it will be observed that unlike for the "unipolar" self-gravitating vortices, where the environment is not a governing factor, for the dipole toroidal vortex, according to (2.4), the environment – similar to vortices in an incompressible fluid – is required in principle. At the same time, a flow generated "lifting force" can explain the existence of "thick" cold [3] tori that causes per se a problem now [18].

In the luminosity, let us single out the contribution to the accretion disk of torus matter and the "background" luminosity $L_0$ unrelated to torus:

$$L = L_0 + \xi \dot{M} c^2, \qquad (2.5)$$

---

[3] The latter is necessary for the existence of dust and is immediately confirmed by the IR observation [4]

where $\dot{M} \equiv dM/dt$ is the accretion rate[4], and $\xi \sim 0.1$ means accretion related energy conversion into radiation. The magnitude $L$ will be considered close to the Eddington limit, which is typical of the AGN luminosity. The toroidal vortex luminosity contribution is described by the second addend connected with the accretion disk vortex "twisting".

### 3. Vortex replenishment of an accretion disk

For the problem considered, the vortex matter inflow into an accretion disk due to particle detachment in the region of contact of dipole components is essential. This process is similar to that considered in [13] of the jet origin in a compact-vortex hole (Fig. 4).

The said process will be described phenomenologically by introducing the effective "height" $h$ of a belt through which the toroidal vortex matter flows into a disk. Then the mass flow towards the disk per unit time is equal to

$$\dot{M} = \rho v_\varphi \cdot 2\pi R \cdot h, \qquad (3.1)$$

where the vortex density $\rho$ is

$$\rho \equiv m_H n = \frac{M_{ring}}{2\pi R \cdot \pi r^2}, \qquad (3.2)$$

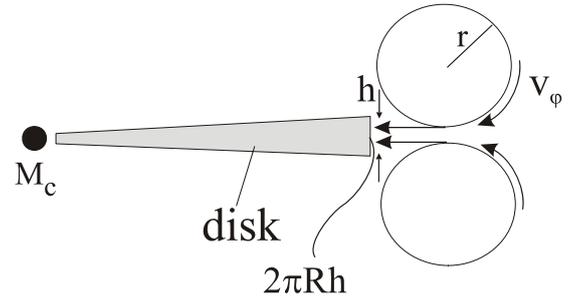

Fig. 4. Scheme of a vortex-fed accretion disk. The belt effective height h is amenable to particle intake in a disk.

and the vortex velocity $v_\varphi$ is expressed through circulation $\Gamma$ as

$$v_\varphi = \frac{\Gamma}{2\pi r}. \qquad (3.3)$$

The particle detachment parameters enter into the expression for the area $2\pi R \cdot h$ through the belt height which will be taken equal to the torus minor radius part

$$h = \xi_1 r. \qquad (3.4)$$

The major and minor radii relate as [13]

$$r = \sqrt{\lambda R}, \qquad (3.5)$$

where $\lambda = \Gamma^2/(4\pi G M_{ring})$ is the Jeans scale. This relation is a direct consequence of coordinate dependencies of gravitational and centrifugal forces connected to a vortical motion in a torus. If the

---

[4] If matter enters the accretion disk only from a torus, then in the symmetrical case $\dot{M} = -2\dot{M}_{ring}$

thermal pressure should be taken into account, the relation [13] will be used

$$r = \frac{\Gamma/2\pi}{\sqrt{\frac{GM}{\pi R} - \frac{1}{2}c_s^2 j_1^2}}, \qquad (3.6)$$

where $c_s$ is the sound speed, $j_1$ is the geometric quotient of about unit. By substituting (3.2) - (3.5) in (3.1) we obtain the accretion rate expression

$$\dot{M} = \frac{2\xi_1 G}{\pi} \frac{M_{ring}^2}{\Gamma R}. \qquad (3.7)$$

The magnitude $\dot{M}$ determines the accretion rate defining the central source luminosity and connected to the replenishment from a toroidal vortex.

Generally speaking, by virtue of nonstationarity of the process under investigation, the time delay between the mass intake into a disk at the distance of a torus major radius $R$ and its "irradiation" in the central engine (i.e. in the disk inner part) may become essential. The effect of this irradiation, as well as of the time delay between moment of radiation and vortex twisting (due to the light and wind speed finiteness), will be discussed more below.

Now let us substitute the accretion rate expression (3.7) into the luminosity formula (2.5)

$$L = L_0 + \frac{2\xi\xi_1 Gc^2}{\pi} \frac{M_{ring}^2}{\Gamma R}. \qquad (3.8)$$

Hence, using the relation of $\Gamma$ with $p_\varphi$ and taking (3.5) into account yield the following formula for the rate of momentum change (2.1) due to twisting:

$$\left(\frac{dp_\varphi}{dt}\right)_{twist} = \frac{\pi^2 \zeta(\theta) L_0}{2GM_{ring}^3 c} p_\varphi^2 + \frac{\zeta(\theta)\xi\xi_1 c}{2R} p_\varphi. \qquad (3.9)$$

The torus mass loss during replenishing the disk results, however, in the loss of the momentum carried away by the escaping (pulled inward a disk) mass. The corresponding momentum losses are described by

$$\left(\frac{dp_\varphi}{dt}\right)_{repl} = -\frac{\xi_1 G}{\pi^2 R} M_{ring}^2. \qquad (3.10)$$

Actually, the momentum carried away from a torus per unit time is equal to

$$\left(\frac{dp_\varphi}{dt}\right)_{repl} = -\rho v_\varphi \cdot r v_\varphi \cdot 2\pi R h, \qquad (3.11)$$

whence follows (3.10).

The rate of luminosity change is determined from (3.8)

$$\frac{dL}{dt} = \frac{2\xi\xi_1 Gc^2}{\pi} \frac{d}{dt} \frac{M_{ring}^2}{\Gamma R}.$$ (3.12)

Being interested in rather fast rates, we may consider the major radius $R$ as slowly varying and substitute $\dot{M}_{ring}$ using the expression (3.7). Then

$$\frac{dL}{dt} = \frac{2\xi\xi_1 Gc^2}{\pi} \frac{M_{ring}^2}{(\Gamma R)^2} \left[ \frac{4\xi_1 GM_{ring}}{\pi} - \frac{d}{dt}(\Gamma R) \right].$$ (3.13)

Time evolution of the inequality $dL/dt > 0$ depends essentially on how the mass $M_{ring}$ and torus major radius $R$ change. If the torus illuminated side is exposed to wind and radiation, then on its shady side there is no radiation pressure and the mass inflow is possible from a more distant environment. In particular, one of the variants of the discussed scenario corresponds to the steady-state mass inflow which allows to consider $\dot{M}_{ring}$ as one of the slowly varying parameters for the times of "fast" variations.

### 4. Accretion-wind instability

Let us first neglect the losses, assuming that the inequality which provides the angular momentum growth is fulfilled:

$$\left(\frac{dp_\varphi}{dt}\right)_{twist} > -\left(\frac{dp_\varphi}{dt}\right)_{repl}.$$ (4.1)

Substitution of (3.8) at $L_0 = 0$ and (3.9) in (4.1) yields momentum restriction from below

$$p_\varphi > \frac{2GM_{ring}^2}{\pi^2 \xi\varsigma(\theta)c}.$$ (4.2)

The angular momentum which satisfies (4.2) on the order of magnitude is equal to

$$p_\varphi^* \approx \frac{GM_{ring}^2}{\xi\varsigma(\theta)c}.$$ (4.3)

This might be amenable to the fact that the contribution of the "background" addend with $L_0 \neq 0$ into vortex twisting can be of fundamental importance [6]. The accretion rate and luminosity magnitudes corresponding to $p_\varphi^*$ are of the form

$$\dot{M}^* = \frac{\xi\xi_1\varsigma(\theta)M_{ring}c}{\pi^2 R} , \quad L^* = \frac{\xi^2\xi_1\varsigma(\theta)M_{ring}c^3}{\pi^2 R}.$$ (4.4)

The possibility for AGN instability connected both with the accretion from a torus and with the central source wind (photon wind included) is obvious from (3.8). The nature of such accretion-wind instability, as it could possibly be named, is that the growing $L$ increases $\dot{p}_\varphi$, while this in turn increases $\dot{M}$, that again increases $L$. Linear increment of the accretion-wind instability at $L_0 = 0$ should result in the exponential growth with the slowly rising (due to the decrease of $R$) parameter $\alpha$:

$$\frac{dp_\varphi}{dt} = \alpha \cdot p_\varphi, \quad \alpha = \frac{\xi \xi_1 \varsigma(\theta) c}{2R}. \tag{4.5}$$

The complete analysis of the instability may appear rather complicated and is not meant here in this paper. Nevertheless, in its character and behavior the instability is similar to the observable quasar radiation bursts [19] that can testify to the discussed dynamic role of the AGN toroidal vortices (see Fig. 7 and discussion further in this paper).

## 5. The delay effect on the increment

The feedforward and feedback circuit, which generates the accretion-wind instability, has the delay which in the oversimplified case is described by the equation

$$\frac{dp_\varphi(t)}{dt} = \alpha p_\varphi(t - \tau), \tag{5.1}$$

where $\tau = \tau_1 + \tau_2$ (see Fig. 5). The evolutionary differential equation with the time delay

$$\frac{dx(t)}{dt} = \alpha x(t - \tau) \tag{5.2}$$

allows the exact solution and results, as it was earlier in the system with no delay, in the exponentially growing solution

$$x(t) = x(0) \cdot e^{\alpha f(\alpha \tau) \cdot t}, \tag{5.3}$$

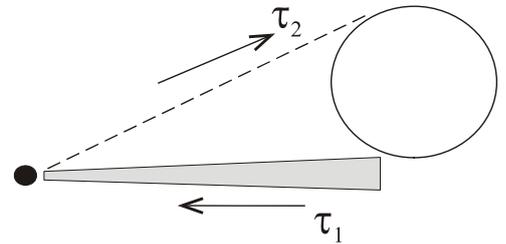

Fig. 5. Scheme of the time delay in a feedback circuit of accretion-wind instability, where $\tau_1$ is the time of mass transfer along the disk radius, $\tau_2$ is the propagation time of centre-to-torus radiation.

though with the increment depending on the dimensionless delay $\alpha \tau$, where the universal function $f(\alpha \tau)$ is the solution of the transcendental equation $f = \exp(-\alpha \tau f)$ or

$$\frac{\ln f}{f} = -\alpha \tau. \tag{5.4}$$

The form of this function is shown in Fig. 6 and is obtained through inversion of the function (5.4).

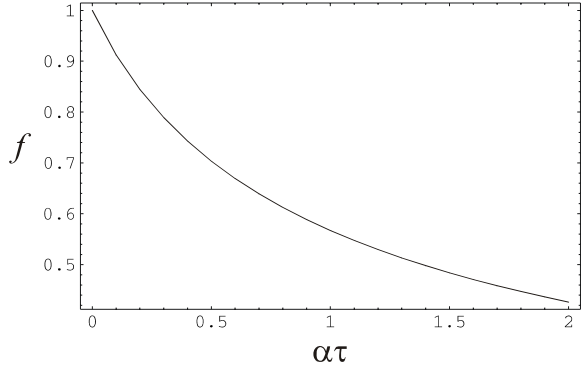

Fig.6. The dependence $f(\alpha\tau)$ obtained as graphic solution of the functional equation (5.4).

It is seen that $f \leq 1$, $f = 1/(1+\alpha\tau)$ for the weak delay $\alpha\tau \ll 1$. Thus, the delay occurrence reduces the increment of accretion-wind instability $\alpha \rightarrow \alpha \cdot f(\alpha\tau)$. With the vortex compression and thin-to-compact evolution [13], the delay becomes less important and the increment rises.

It is interesting that the equilibrium condition for a thin vortex with slow orbital motion $\lambda_\theta \ll \lambda_\varphi$, where $\lambda_{\theta,\varphi} \equiv \pi p_{\theta,\varphi}^2 / GM^3$ also reduces to the equation of a form (5.4):

$$\beta^2 \frac{\lambda_\varphi}{R} = e^{-\lambda_\theta/R}. \qquad (5.5)$$

Here $\beta \sim 1$ is the parameter included into the potential energy of a self-gravitating toroidal vortex [13] $U(r,R) = G\frac{M^2}{\pi R}\ln\frac{\beta r}{R}$. Hence, for $r \rightarrow R$, a slow orbital motion influences on the inner radius of a compact torus $R - r(R) \approx \lambda_\theta$, while the outer radius $R \approx \lambda_\varphi$ is determined by the vortex motion.

### 6. Discussion

The aforesaid counts in favour of the considered system to possibly having the positive feedback and the instability it generates. The origin of accretion-wind instability can be explained as simple as follows: it is obvious that $\dot{p}_\varphi \propto L$, $L \propto \dot{M}$. As follows from the expression for $\dot{M}$ (3.1), we may suppose $\dot{M} \propto p_\varphi$, that closes the positive feedback loop. In reality, the situation is rather more complicated as the loop closure uses the dependence of the minor radius of a vortex vs. its major radius (3.5) through the moment dependent Jeans scale. Moreover, it appears that $\dot{M} \propto 1/p_\varphi$ (3.7). This, however, does not alter the situation with the feedback, as $\dot{p}_\varphi$, according to (3.9), is proportional not merely to $L$ but to $Lp_\varphi^2$ that results in equation (4.5).

We have considered the instability in its simplest case with least parameters: no background radiation except that connected with a toroidal vortex, insignificant wind vs. radiation contribution to twisting, rather slow motion (contraction) over the torus major radius. Then, according to (4.5), the accretion-wind instability increment is equal to $\alpha = \xi\xi_1\varsigma(\theta)c/(2R)$ and determines the characteristic time scale of its development. In case we want to compare a 3.5-year time scale of the observable burst duration in the quasar 3C 345, see Fig. 7 from [19], we should take the space scale $R \sim 10^{16}$ cm (1/300 light year). Here we have taken into account that $\xi \sim 0.1$, and $\xi_1$ is taken of the same infinitesimal order. The last estimation may essentially differ from reality, therefore $\xi_1$ should be sooner treated as a scale factor which variation may considerably change the system parameters.

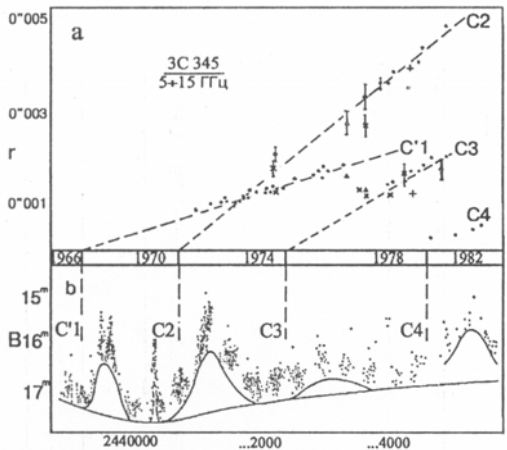

Fig. 7. Correlation between the optical bursts of the quasar 3C345 and arising the super luminal components of radio jet [19].

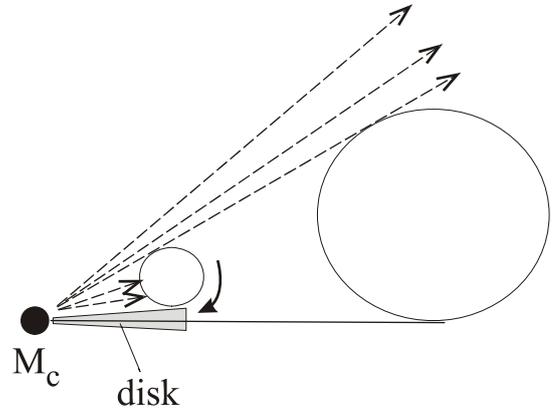

Fig. 8. The possible obscuring AGN structure in the form of the "Matreshka" dolls sequence of tori.

As the preliminary observed data [4] point to significantly larger torus sizes, a question – which one of the answers results in the "Matreshka" scheme (Fig. 8) – may naturally arise. The inner toroidal vortex may be responsible for the AGN variability, the development of instability, etc. In the shadow of a nearest-to-the-center smallest-radius torus there exist a preference for the center-falling matter because of least-interfering radiation (this latter being weaker due to absorption) and of wind screened by the inner torus. Therefore the interstellar gas clouds will move towards the center in the nearby torus shadow. Orbital motion causes the falling clouds to flatten into tori and disks. Though the outer tori cannot add to developing the accretion-wind instability owing to this latter extremely slow development at large scales (cf. the estimate (4.5)) and, in addition, being weakened by the increment-slowing delay. (The properties of self-gravitating tori as attractors see in Appendix B.)

Thus, the instability is determined by the inner center-closest torus. The torus distribution increment is determined by the accretion process on the scales significantly exceeding those considered here and is beyond the discussed scenario. Detailed studying of the processes which occur in the evolution of toroidal vortices in the centers of active galaxies is a highly intricate problem. Nevertheless, it is possible already now to distinguish some features of these processes. Under the Eddington luminosity close conditions, due to the radiation pressure compensated center attraction, the evolution of vortices should largely resemble their evolution without the central mass [13]. At the compact vortex phase, the ejection of particles along the torus axis is possible, that might explain the correlation between the quasar optical bursts and the formation of new jet components [19] (see also discussion of correlation problem and references for example in [20, 21]).

The expressions obtained above may allow to estimate the features of the outer (obscuring) torus for the Seyfert galaxies (see Table 1) and the quasars (see Table 2).

Table 1. Parameters of the obscuring torus for Seyfert galaxies.

| Model parameters | Chosen values | Calculated AGN values | Obtained values |
|---|---|---|---|
| $M_{BH}$ | $6.6 \cdot 10^7 M_\odot$  [22] | $\alpha$ | $5 \cdot 10^{-12}$ CGS |
| $M_{ring}$ | $0.1 \cdot M_{BH}$  [22] | $p_\varphi$ | $3.8 \cdot 10^{64}$ CGS |
| R | 1pc  (see [3]) | $\Gamma$ | $1.8 \cdot 10^{25}$ CGS |
| r/R | 0.5  (see [23]) | $v_\varphi$ | $2 \cdot 10^6$ cm/s |
| $\xi$ | 0.1 | $n$ | $5.4 \cdot 10^7$ cm$^{-3}$ |
| $\xi_1$ ; $\varsigma$ | 0.1 | $nV_{ring}^2$ | $2 \cdot 10^{20}$ CGS |
| $n_{wind} U_{wind}^2$ | $10^{22}$ CGS ($n_{wind} = 10^6$ cm$^{-3}$, $U_{wind} = 10^8$ cm/s) | $nV_{esc}^2$ | $7.7 \cdot 10^{21}$ CGS |
| $L_{Edd}$ | $8.6 \cdot 10^{45}$ erg/s | $L^*$ | $1.2 \cdot 10^{48}$ erg/s (see the text) |

Table 2. Parameters of the obscuring torus for quasars.

| Model parameters | Chosen values | Calculated AGN values | Obtained values |
|---|---|---|---|
| $M_{BH}$ | $10^9 M_\odot$ | $\alpha$ | $5 \cdot 10^{-12}$ CGS |
| $M_{ring}$ | $0.1 \cdot M_{BH}$ | $p_\varphi$ | $8.8 \cdot 10^{66}$ CGS |
| R | 1pc | $\Gamma$ | $2.8 \cdot 10^{26}$ CGS |
| r/R | 0.5 | $v_\varphi$ | $3 \cdot 10^7$ cm/s |
| $\xi$ | 0.1 | $N$ | $1.4 \cdot 10^9$ cm$^{-3}$ |
| $\xi_1$; $\varsigma$ | 0.1 | $nV_{ring}^2$ | $6.7 \cdot 10^{23}$ CGS |
| $n_{wind}U_{wind}^2$ | $5 \cdot 10^{24}$ CGS, [24] | $nV_{esc}^2$ | $1.8 \cdot 10^{24}$ CGS |
| $L_{Edd}$ | $1.3 \cdot 10^{47}$ erg/s | $L^*$ | $1.8 \cdot 10^{49}$ erg/s (see the text) |

Note that in [22], the dust mass of an obscuring torus is estimated on the order of magnitude of $0.01 M_{BH}$, where $M_{BH}$ is the central black hole mass. In our estimations, we assume dust making 10% of the total torus mass.

The discrepancy between the characteristic $L^*$ and the Eddington luminosities $L_{Edd}$ can be easily eliminated by assuming a smaller efficiency replenishment of the accretion disk by a toroidal vortex. Thus, taking $\xi_1 = 10^{-3}$ yields $L^* \sim L_{Edd} \sim 10^{47}$ erg/s for the other parameters unchanged. However, the delay-driven decrease of $L^*$ may appear to be essential as well. The luminosity $L^*$ (4.4) can be represented as

$$L^* = 2 \xi \alpha M_{ring} c^2 / \pi^2 , \qquad (6.1)$$

where $\alpha$ is the accretion-wind instability increment. As is shown above (see item 5), the $\tau$-time delay decreases the increment by a factor of $f$, where $f(\alpha \tau)$ is the solution of equation (5.4). As a matter of fact, the time delay essentiality means that the detail description needs using the theory of a nonstationary disk accretion with the "boundary conditions" determined by the interaction of a toroidal dipole vortex with a disk, that exceeds the bounds of this paper.

For the luminosity close to the Eddington limit, the torus mass $M_{ring}$ have to be near to the value $M_{ring} = \eta(R) M_c$, where $\eta(R)$ is estimated from the relation $L^* < L_{Edd} = 1.3 \cdot 10^{38} \left( \frac{M_c}{M_\odot} \right)$ erg/s, that led to inequality $\eta(R) < 3.7 \cdot 10^{-6} \frac{1}{\xi_1 \varsigma(\theta)} \left( \frac{R}{1 pc} \right)$. In particular, in the case $R = 10^{-2} pc$ for $M_c = 10^9 M_\odot$ and $\xi_1 = \varsigma = 0.1$ we obtain the value $M_{ring} \approx 4 \times 10^3 M_\odot$. The decrease of torus mass with the decreasing torus radius can naturally be connected with mass departure to the accretion disk and/or with blowing off some part of the mass under the action of the wind.

The torus twisting by wind rather than by radiation may appear essential. In this case, the magnitude $\rho_{wind} U_{wind}^2$ will play the role of pressure on a torus in (2.1). Closing a feedback circuit requires the knowledge about the connection between wind parameters and the central source luminosity. The magnetic field which impacts the parameters of wind and its angular distribution, and accordingly the torus twisting, can be of importance too[5]. Despite of these problems yet unsolved, the described scheme already now yields the reasonable correspondence with the data observed until recently.

A short description of some items of this work is published in the authors' paper [29].

## 6. Conclusion

1) A dipole toroidal vortex may be an essential AGN-structure element which "replenishes" the accretion disk.

2) In the feedback circuit, which includes vortex twisting by radiation and wind, and vortex replenishment of the accretion disk, the instability causing the bursts in active nuclei may develop.

3) The presence of a centrifugal force in the toroidal vortex and a "lifting" force due to wind flow may allow the existence of a "thick" and cold torus.

4) The "Matreshka" scheme of an AGN toroidal structure which may explain the evolution and variability effects is proposed.

---

[5] A torus vortex motion can be caused by the magneto-rotational instability owing to the presence and constant amplification of the toroidal field in the disk (see [25]). In this case, the origin itself of the vortex torus can be related to instability in the disk. (Authors are thankful to the reviewer for this remark.) The magnetic field may be responsible also for existing cold thick tori [26]. Generally speaking, the impact of even a weak magnetic field and its topology can be essential in shaping the toroidal structures. The MHD simulation example [27] – corresponding, however, not to accretion but to ejection – gives rather an accurate account of what we propose, though with the opposite current direction in a disk and rotation in a torus. A modern review of magnetic field influence on flows in AGN central regions see monographs [28].

## Appendix A. Bernoulli effect in the dipole self-gravitating vortex flow.

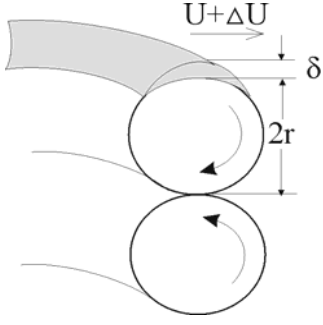

Fig. 9. Torus surface perturbation (to the wind lift calculation).

The work done by the wind flow of two components of a dipole vortex can be noted as $A_{wind} = 2 \cdot \Delta p \Delta V$, where $\Delta p = \rho_{wind} U \Delta U$; $\Delta p$ and $\Delta V$ are the variations of pressure and volume for the torus displacement by the magnitude $\delta$, respectively (see Fig. 9). With the $\Delta U \approx U \frac{\delta}{r}$ taken into consideration, we obtain:

$$A_{wind} = 2\rho_{wind} U^2 \frac{\delta}{r} \Delta V. \qquad (A.1)$$

The gravitational attraction between the two rings with masses $M_a$ and $M_b$, with taking into account the distance between them being equal to 2r, has the form [30]:

$$F_g = \frac{GM_a M_b \cdot (2r) \cdot k}{4\pi R^3} \frac{E(k)}{1-k^2}, \qquad (A.2)$$

where $k = R/\sqrt{r^2 + R^2}$, $E(k)$ is the elliptic function. If $r \ll R$, then $k \approx 1 - 1/2 \cdot (r/R)^2$ and $1 - k^2 \approx (r/R)^2$. In this case $E(k) \approx E(1) = 1$ and, therefore, the gravitational attraction between the two components of a dipole toroidal vortex takes the form

$$F_g = \frac{GM_a M_b}{2\pi Rr}. \qquad (A.3)$$

The work $A_g$ done by gravitational attraction of the region of elevation $\delta$ to the dipole vortex is equal to $F_g \cdot \delta$. With the $M_a = M_{ring} \delta/r$, $M_b = 2M_{ring}$, and the volume change $\Delta V = 2\pi R r \cdot \delta$ considered in (A.3), we obtain

$$A_g = \frac{2GM_{ring}^2}{(2\pi Rr)^2} \cdot \frac{\delta}{r} \cdot \Delta V. \qquad (A.4)$$

Comparing (A.1) and (A.4), we arrive at the sought expression

$$\rho_{wind} U^2 = \rho_{ring} V_{esc}^2, \qquad (A.5)$$

where $V_{esc}^2 = GM_{ring}/(2R)$ is the characteristic escape velocity.

## Appendix B. Particle-to-torus attraction

Using the simple reasoning may show that a test particle near the inner side of a self-gravitating torus is subjected to the force attracting it to a torus. First, let us consider a thin torus which extreme case is a cylinder. In fact, the test particle will be attracted to the cylinder. This alone means that the test particle inside a torus is subjected to the force directed to this latter, or to put it more precisely – to the torus part nearest to the particle.

Let us represent a torus as a system of concentric rings. Consider the elementary case, i.e. attraction of a test particle to diametrically opposite ring areas. Select two "sectors", with their vertices at the test particle and with a small angular span, symmetric with respect to diameter of the ring passing through the test particle. Let us consider the forces of particle attraction to the opposite arcs of a ring inside sectors. If a particle is at the centre of the ring, they balance each other (Fig. 10a). When decentering the particle (along the chosen diameter), we may see that the arc mass increases (or decreases) linearly with distance from the particle, while the force changes inversely as square of distance (Fig. 10b). Therefore, the attraction force from a "distant" arc decreases, though the arc length increases, while the force of attraction to the nearest arc increases, though the arc length decreases with the particle approaching it.

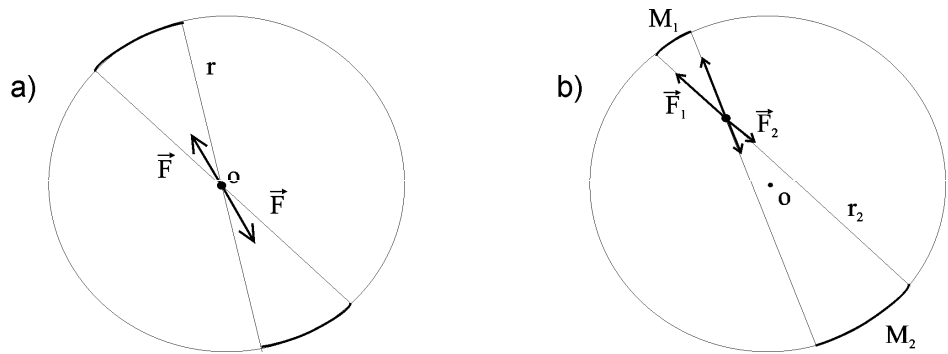

Fig. 10. Scheme of a test particle-to-ring attraction

An uncompensated force of particle-to-ring attraction and, accordingly, that of particle-to-torus appear. Expanding the span angle we are compelled to proceed from elementary formulas to integrals, though this does not change anyhow the fact of the matter and the result. It will be noted that in the case of a sphere (with the similar reasoning) the mass attracting a particle is proportional to the area on a sphere cut by a solid angle. Therefore, decentering the particle saves the exact compensation of forces: the mass is changed quadratically with distance and is compensated by inverse dependence of the force vs. squared distance. Therefore, as is notorious, a test particle inside a sphere (as against a torus) is subjected to no gravitational force.

The previous reasoning is sustained by the calculation of trajectories of test particles (Fig. 11). Examples of trajectories for the "vertical" symmetry plane motion are shown in [13]. As can be seen from Fig. 11a and Fig. 11b, at small energies, the particle travels around a circle with minor radius, coiling around a ring (Fig. 11a shows the particle planar motion, Fig. 11b shows the presence of orbital motion). Such motions correspond to a thin vortex (the first stage of evolution possible). With larger particle energy, different complicated trajectories appear (Fig. 11c). And finally, beginning from some amount of energy, they transform into the almost closed figure-of-eight loop type trajectories (Fig. 11d). At the same time the rotation radius of particles becomes about that of a ring which is characteristic of a compact phase of vortex compression. The kinetic energy of a particle at such trajectories is already close to the energy sufficient for the particles to escape. The figure-of-eight loop motion of particles will result in their effective collision and scattering near the vortex axis, which, as is noted in [13], is capable to result in the origin of an unidirectional jet along the axis. The existence of a figure-of-eight type flows of matter follows from the Maxwell's hydrodynamic vortex model, too (see, e.g. [15]).

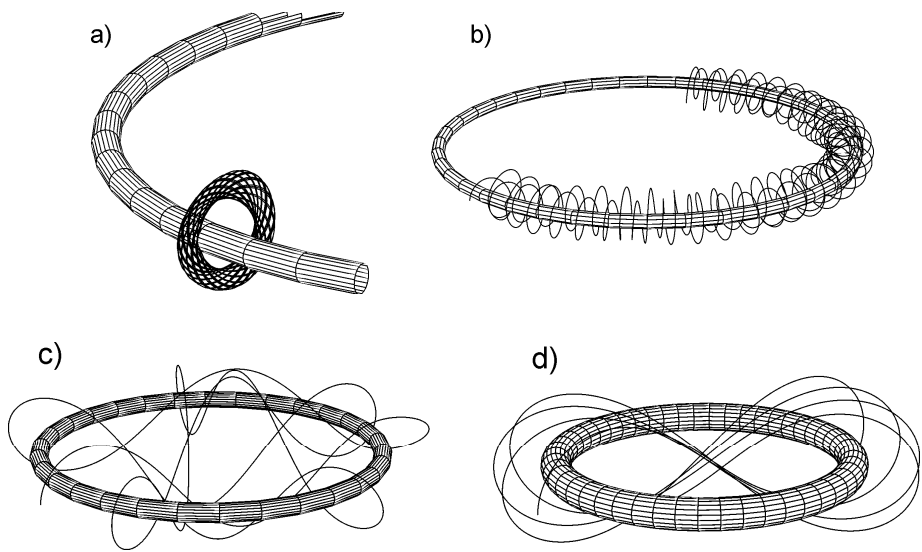

Fig. 11. 3D finite trajectories of a test particle in the gravitational field of a ring.

*This version of paper was translated by A.E. Kirichenko*